\documentclass[amssymb,aps,prl,showpacs,superscriptaddress]{revtex4}
\usepackage{amsmath}
\def\be{\begin{equation}}
\def\ee{\end{equation}}
\def\Tr{{\rm Tr}}
\def\e{{\rm e}}
\def\dif{{\rm d}\,}

\begin{document}

\title{Timescales for decoherence and dissipation: where does the success 
of master equations come from?}
\author{M.\ O.\ Terra Cunha\thanks{%
tcunha@mat.ufmg.br}}
\affiliation{Departamento de Matem\'atica, Universidade Federal de Minas Gerais, C.P. 702,
Belo Horizonte, MG, 30123-970, Brazil}
\affiliation{Departamento de F\'{\i}sica, Universidade Federal de Minas Gerais, C.P. 702,
Belo Horizonte, MG, 30123-970, Brazil}
\author{S. Geraij Mokarzel}
\affiliation{Instituto de F\'{\i}sica, Universidade de S\~ao Paulo, C.P. 20516,
S\~ao Paulo, SP, 01317-970, Brazil}
\author{J.\ G.\ Peixoto de Faria}
\affiliation{Departamento de Ci\^{e}ncias Exatas e Tecnol\'{o}gicas,
Universidade Estadual de Santa Cruz, Ilh\'{e}us, BA, 45662-000, Brazil}
\author{M.\ C.\ Nemes}
\affiliation{Departamento de F\'{\i}sica, Universidade Federal de Minas Gerais, C.P. 702,
Belo Horizonte, MG, 30123-970, Brazil}
\date{\today}

\begin{abstract}
A perturbative treatment of reduced density operators of quantum subsystems
is implemented in the same spirit as Fermi Golden Rule for scattering.
Analytic expressions for linear entropy (a measure of purity loss, and in some
cases of coherence loss) and for subsystem's energy variations (dissipation)
are given. Application to electromagnetic field superposition states in a
dissipative cavity is performed. Evaluation of typical field and reservoir
time scales ($\tau _{F}$ and $\tau _{R}$) show that even for small
temperatures they are very different. We also indicate the condition on the cutoff temperature 
above which the decoupling assumption is quantitatively justified. 
\end{abstract}

\pacs{03.65.-w, 03.65.Bz, 03.65.Ca  }
\maketitle

Since 1970 many techniques for controlling single quantum systems have been developed, 
\textit{e.g.}, trapping a single ion in what is called an ``ion trap''
\cite{ionTraps}, or storing a single mode of electromagnetic field in
high quality cavities \cite{cavity},
trying to isolate them from the rest of 
the world and thus allowing us to probe many different aspects of its behavior with 
amazing precision. However, the control over properties of quantum systems, the 
essential ingredient  for the technological implementation of quantum computers 
(as \textit{e.g.} entanglement among many qubits), is strongly hindered by deleterious 
environmental effects. The first experiment which clearly exhibits the effect 
of the environmental dynamics on a mesoscopic quantum superposition has been 
performed at the Laboratoire Kastler-Brossel in Paris \cite{paris}.

From the theoretical point of view several independent ideas have been developed to 
treat open quantum systems \cite{openQuSys}. The best known and used is the approach with 
master equations. Its starting point consists of a system of interest ($S$) 
linearly coupled to a 
``reservoir'' ($R$) usually simulated by a large number of harmonic oscillators 
(for a complete derivation see \cite{mod, cohen, Mte}). The trace 
over the reservoir degrees of freedom leads to an effective equation for the system 
of interest containing nonunitary operators, responsible for coherence and energy loss, where
the reservoir appears only through some mean values, just like in Thermodynamics.
Several approximations are necessary in order to obtain such linear, markovian equation 
for $S$. One approximation which needs physical support to rest on \cite{Amir}
is the one which considers the reservoir density
uncoupled from $S$ \emph{for all times}\footnote{It is usual to state only that
there are no correlation between $R$ and $S$ at the initial time, use it to
derive the master equation, and then apply it to any subsequent time. This
is equivalent to remake the same factorizability assumption at any time.}. It
can not be rigorously true, since decoherence of $S$ system can only come from
correlations with $R$. As discussed by Cohen-Tannoudji and co-workers \cite{cohen},
this is easy to accept it provided the temperature is large enough so that the typical time 
scales are very different. This tends to make any previous correlation between $R$ and $S$
irrelevant after a short time. However, we know from Schmidt's decomposition
that if we have
any pure bipartite system, the evolution of correlations is exactly the same for both 
subsystems. In this case, a ``cold'' reservoir consisting of all oscillators in the 
ground state belongs to this category and therefore the argument breaks down. In spite 
of this fact, master equations usually work extremely well also for zero temperature
\cite{lowTemp}. Why? This is one of the long standing questions in the area 
of open quantum systems.

The objective of the present contribution is to work out the typical time scales 
for variation of coherence and energy in any two coupled quantum systems. When 
one of the systems is a reservoir we obtain a limiting temperature for the 
validity of the factorization assumption holds.
For this purpose we next implement a perturbative treatment of reduced matrices 
of quantum systems in the same spirit as the Golden Rule for scattering.

Consider a Dyson like expansion for reduced operators which obey the following Hamiltonian dynamics
\be
H = H_{1} + H_{2} + H_C \equiv H_0 + H_C
\ee
where $H_0$ contains the autonomous evolution of the systems in question and $H_C$ 
represents the coupling between them. We will obtain perturbative equations for the reduced densities 
\be 
\rho_i = \Tr_j\rho(t)\,,
\ee 
where $\rho(t)$ is the total density, $i=1,2$, $j=1,2$ and $i\neq j$ are subsystem's indexes. We get
\be\begin{split}
\rho_i(t) & = \rho_i(0) - i\int_0^t\dif t' \Tr_j\bigl[H_C(t'),\rho(0)\bigr] \\
& \qquad -\int_0^t\dif t' \int_0^{t'}\dif t'' \Tr_j\bigl[H_C(t'),\bigl[H_C(t''),\rho(0)\bigr]\bigr] + 
\dotsb\quad (i\neq j) \\
& \equiv \rho_i^{(0)} + \rho_i^{(1)} +\rho_i^{(2)} + \dotsb\,,
\label{born}
\end{split}\ee
where $H_C(t) \equiv \e^{iH_0t}H_C\e^{-iH_0t}$ (the interaction representation is applied). Note the symmetry in the indexes $i,j$, 
which will allow us to study the short time dynamics of any of the two subsystems. Instead 
of proceeding formally and rewriting the above equation in terms of matrix elements, we prefer 
to illustrate the procedure for obtaining dissipation and decoherence time scales by mean of 
an example. Let ($\hbar = 1$)
\be
H_0 \equiv \omega a^\dag a + \sum_j \Omega_j b_j^\dag b_j,
\ee
which represents, \textit{e.g.}, the free Hamiltonian of a given cavity mode plus that of the environment 
simulated by oscillators as usual. $\omega$, $\Omega_j$, $a(a^\dag)$ and $b_j(b_j^\dag)$ 
are the natural frequencies, annihilation (creation) operators of the mode and the reservoir, 
respectively. We assume that the coupling between the two systems is given by
\be
H_C = \sum_j \gamma_j \bigl(a^\dag b_j + ab_j^\dag\bigr)\,,
\ee
where $\gamma_j$ are the coupling constants assumed to be real.

\emph{Subsystems dissipation timescales}: Let us define a subsystem energy by 
\be
E_i(t) = \Tr \left[H_{i}\rho_{i}(t)\right]
\label{eq5}
\ee
and assume that we initially have the state of the composed system given by
\be
\rho(0) = \rho_1(0)\otimes\frac{\e^{-\beta H_{2}}}{\Tr\bigl[\e^{-\beta H_{2}}\bigr]}\,,
\label{initial}
\ee
where $\beta = 1/(k_B T)$. Inserting Eq.~\eqref{born} in \eqref{eq5} we get, up 
to second order for the cavity mode labeled 1, 
\be
E_1(t) = E_1^{(0)} + E_1^{(1)}(t) + E_1^{(2)}(t) + \dotsb\,,
\ee
where the upper indexes indicate the order of the approximation. We get
\begin{align}
E_1^{(0)} & = \Tr_1\bigl[\omega a^\dag a \rho_1(0)\bigr] = 
\omega \langle a^\dag a\rangle_{0} = E_1(0), 
\label{E0th} \\
E_1^{(1)}(t) & = -i\omega\int_0^t\dif t' 
\Tr\bigl\{a^\dag a[H_C(t'),\rho(0)]\bigr\} = 0, 
\label{E1st} \\
E_1^{(2)}(t) & = -\omega \int_0^t\dif t'\int_0^{t'}\dif t'' 
\Tr\bigl\{a^\dag a[H_C(t'),[H_c(t''),\rho(0)]]\bigr\} \nonumber \\
& =  \omega\sum_j \left(\frac{\gamma_j \sin\Delta_j t}{\Delta_j}\right)^2
\bigl(\bar n_j - \langle a^\dag a\rangle_{0}\bigr)\, ,
\label{E2nd} 
\end{align}
where $\Delta_j = \left(\Omega_j-\omega\right)/2$, 
$\bar n_j = \left(e^{\beta \Omega_j} - 1\right)^{-1}$ is the average 
occupation number of the 
$j$-th reservoir state at temperature $T$, and $\langle a^\dag a\rangle_{0} $
is the average initial excitation number of the subsystem 1. 
Note in the above equation the basic ingredients 
of Fermi's Golden Rule: each reservoir mode contributes to the system energy initially in a way proportional to its coupling constant and to the difference in occupation numbers between the mode and the system. As time goes by, the
$\left( \sin\Delta_j t/\Delta_j\right) ^2$ function selects only those modes with frequency very close to $\omega$. In scattering language, this is the ``phase space'' available for the process.

In order to gain deeper physical insight into the energy and coherence loss
processes, let us consider the subsystem 1 initially in the so called even 
coherent state \cite{DMM74} with amplitude $\alpha $, 
$\rho _{1}\left(0\right)=\left| e\alpha \right\rangle \left\langle e\alpha \right|$, where 
\[
|e\alpha\rangle = \frac1{2\bigl(1+\e^{-|\alpha|^2}\bigr)}
\bigl[|\alpha\rangle + |-\alpha\rangle\bigr]\,.
\]
Here, $\left| \alpha
\right\rangle $ is Glauber's coherent state \cite{Gla63}.
The average initial excitation number is given by
\be 
\langle a^\dag a \rangle_{0} = |\alpha|^2 \tanh |\alpha|^2\,.
\label{eq:aa}
\ee
To proceed with the calculation we approximate (as usual) the summation over $j$ 
by an integral in Eq. \eqref{E2nd}
\be\begin{split}
\sum_j F_j \left(\frac{\sin\Delta_j t}{\Delta_j}\right)^2 & 
\rightarrow \int_0^\infty\dif\Omega\,g(\Omega)F(\Omega) 
\left(\frac{\sin\Delta t}{\Delta }\right)^2\\
& \approx \int_{\omega-1/(2t)}^{\omega+1/(2t)} \dif\Omega\,g(\Omega)F(\Omega) 
\left(\frac{\sin\Delta t}{\Delta }\right)^2 \approx t^2 \int_{-1/(2t)}^{1/(2t)} 
\dif\Delta\,g(\omega+\Delta)F(\omega+\Delta) = \frac{t}{\tau_F}\,.
\end{split}\ee
where $g(\Omega)$ is the density of modes of the reservoir at frequency 
$\Omega$ and $\Delta = \left(\Omega -\omega\right)/2 $. The first approximation is like a stationary phase argument and the second comes from a linear approximation to the sine function close to $\Delta = 0$. 
This equation defines the time scale $\tau_F$, provided $F$ has frequency squared dimension. 

For the expansion in energy of subsystem 1, we get
\begin{eqnarray}
E_{1}\left( t\right) - E_{1}\left( 0\right) &\simeq& 
\omega t^{2}\int_{0}^{\infty }\mathrm{d}\Omega g\left( \Omega \right) 
\bar{n}\left( \Omega \right) \left[ \gamma \left( \Omega \right) \frac{%
\sin \left( \Delta t\right) }{\Delta t}\right] ^{2} \nonumber \\
&& -\omega t^{2}\left| \alpha
\right| ^{2}\tanh \left| \alpha \right| ^{2}\int_{0}^{\infty }\mathrm{d}%
\Omega g\left( \Omega \right) \left[ \gamma \left( \Omega \right) \frac{\sin
\left( \Delta t\right) }{\Delta t}\right] ^{2} . \label{E2ndcat}
\end{eqnarray}
Note that in the above equation, the second term on the r.h.s. 
is independent of temperature and therefore defines a dissipation time 
scale, whereas the first term, on contrary, depends on temperature and allows 
us to define a thermal timescale. These timescales are determined through the ratio
$\left|E_{1}\left( t\right) -E_{1}\left( 0\right) \right| /E_{1}\left( 0\right) $,
and reads
\be
\tau _{{\rm dis}}^{-1} = t 
\int_{-1/\left( 2t\right) }^{1/\left( 2t\right) }{\rm d}\Delta \
g\left( \omega +\Delta \right) \gamma ^{2}\left( \omega +\Delta \right) , 
\ee
\be
\tau _{{\rm th}}^{-1} = \frac{t}{\left| \alpha \right| ^{2} 
\tanh\left| \alpha \right| ^{2}}  
\int_{-1/\left(2t\right) }^{1/\left( 2t\right) }{\rm d}\Delta \ 
g\left( \omega +\Delta \right) \bar{n} \left(\omega + \Delta\right)
\gamma ^{2}\left( \omega +\Delta \right) .
\ee
In the asymptotic case $t\rightarrow \infty $, 
which corresponds to the Markovian approximation, we obtain 
\be
\tau_\text{dis}^{-1} = g(\omega)\gamma^2(\omega)\,
\label{tauDiss}
\ee
and
\be
\tau_\text{th}^{-1} = \frac{\bar n(\omega)g(\omega)\gamma^2(\omega)}
{\left| \alpha \right| ^{2} \tanh\left| \alpha \right| ^{2}},
\label{tauTh}
\ee
which give us the general timescale relation 
\be
\frac{\tau _{\mathrm{th}}}{\tau _{\mathrm{dis}}} = \frac{\left| \alpha \right|
^{2}\tanh \left| \alpha \right| ^{2}}{\bar{n}\left( \omega \right) }.
\label{relthdis}
\ee
In the large temperature regime $\beta
\omega \ll 1$ we have $\tau _{\mathrm{th}}/\tau _{\mathrm{dis}}
\rightarrow \beta \omega \left| \alpha \right| ^{2}
\tanh \left| \alpha\right| ^{2}$ and Eq. (\ref{relthdis}) has a simple semiclassical
interpretation: as $\left| \alpha \right| ^{2} \tanh \left| \alpha \right| ^{2}$ is proportional to
oscillator's mean energy, the larger $\left| \alpha \right| ^{2}$ is, the
faster dissipation occurs; accordingly, as $\bar{n}\left( \omega \right)
=\exp \left( {-\beta \omega }\right) /\left[ 1-\exp \left( {-\beta \omega }%
\right) \right] $ gives the mean number of quanta of frequency $\omega $ on
thermal reservoir at temperature $T$, the larger $\bar{n}\left( \omega
\right) $, the faster thermal effects appear. Zero temperature is thus characterized by the absence of thermal effects.

\emph{Subsystems decoherence timescales}:
The expression for the idempotency deficit \cite{Ji} of the subsystem $i$, up 
to second order, is given by
\be
\delta _{i}\left( t\right) = 1 - \Tr_{i} \rho_{i}^{2}\left(t\right)
\approx \delta _{i}^{\left( 0\right) }\left( t\right)
+\delta _{i}^{\left( 1\right) }\left( t\right) +\delta _{i}^{\left( 2\right)
}\left( t\right)  , 
\label{expandLinEnt}
\ee
where $\delta _{i}^{\left( 0\right) }\left( t\right) =
\delta _{i}\left( 0\right)$,
$\delta _{i}^{\left( 1\right) }\left( t\right) = -2\mathop{\rm Tr}
\nolimits_{i}\left[ \rho
_{i}\left( 0\right) \rho _{i}^{\left( 1\right) }\left( t\right) \right] ,
\label{deltaS_1}$ and 
$\delta _{i}^{\left( 2\right) }\left( t\right) = 
-\mathop{\rm Tr}\nolimits_{i}\left[ \rho _{i}^{\left( 1\right) }\left(
t\right) \right] ^{2}-2\mathop{\rm Tr}\nolimits_{i}\left[ \rho _{i}\left(
0\right) \rho _{i}^{\left( 2\right) }\left( t\right) \right]$ .
In the last expressions, the operator $\rho _{i}^{\left( n\right) }\left(
t\right) $ represents the  $n$-th term of the expansion  
(\ref{born}). 
For the state given by Eq. (\ref{initial}), we have
$\delta _{1}^{\left( 1\right) }\left( t\right) =0$ and
$\delta _{1}^{\left(2\right) }\left( t\right) = 
-2\mathop{\rm Tr}\nolimits_{1}\left[ \rho
_{1}\left( 0\right) \rho _{1}^{\left( 2\right) }\left( t\right) \right]$. 
Assuming that the subsystem 1 was prepared in the initial state
$\rho _{1}\left( 0\right) =\left| e\alpha \right\rangle \left\langle
e\alpha \right| $, the evaluation of $\delta _{1}^{\left(2\right) }\left( t\right)$ yields,
in the continuum limit, 
\be
\delta _{1}\left( t\right) \simeq 2 \int_{0}^{\infty }\mathrm{d}\Omega g\left(
\Omega \right) \left[ \gamma \left( \Omega \right) \frac{\sin \left( \Delta
t\right) }{\Delta }\right] ^{2}\left\{ \bar{n}\left( \Omega \right) 
\left( \left| \alpha \right| ^{2}\tanh
\left| \alpha \right| ^{2}+1\right)+
\left[ \bar{n}\left( \Omega
\right) +1\right] \left| \alpha \right| ^{2}\tanh \left| \alpha \right| ^{2}
\right\} 
\label{delta2ndcat}. 
\ee
We can also employ the same reasoning that allows us to determinate the
dissipation timescales in order to define the timescale $\tau _{{\rm dec}}$. In the
present case, purity loss and decoherence timescales coincide. Therefore,
within the Markovian approximation, we get 
\be
\tau _{\mathrm{dec}}^{-1}=2 g\left( \omega \right) \gamma ^{2}\left( \omega
\right) \left\{ \bar{n}\left( \omega \right) \left( \left| \alpha \right| ^{2}
\tanh \left| \alpha \right|^{2}+1\right)
+ \left[ \bar{n}\left( \omega \right) +1\right] \left|
\alpha \right| ^{2}\tanh \left| \alpha \right| ^{2} \right\} . 
\label{tauDec}
\ee

As is clear, in the expressions (\ref{E2ndcat}) and (\ref{delta2ndcat}) the
ingredients are the same, but the results are quite distinct. The greater
difference is the relative sign and it is easy to understand it in this
example. In both cases, the first term is the temperature contribution and
the second is still there even for zero temperature. The first term tends to
increase $E_1$ (it is a ``hot'' term), while the second to decrease it (it
is a ``cold'' one), but both add to decoherence. This difference in sign
reflects in the difference in behavior and timescales of these processes.

In case $T=0$ only ``cold'' terms contribute, and we obtain the relation 
\begin{equation}
\frac{\tau _{\mathrm{dis}}}{\tau _{\mathrm{dec}}}=2\left| \alpha \right|
^{2}\tanh \left| \alpha \right| ^{2}  
\label{timescalesT=0}
\end{equation}
and there is no thermalization in the sense which we are using this word. In
the limit $\left| \alpha \right|^{2} \gg 1$,
Eq. (\ref{timescalesT=0}) clearly shows the dependence of the purity loss 
timescale with the separation of the states that form the superposition in
the initial state $\rho_1 \left(0 \right)$. In fact, the more distinguishable 
these states, the smaller the purity loss timescale. 
Such relations of time
scales are usually considered the answer to the impossibility of naturally
finding the great majority of quantum states for macroscopic systems (e.g.:
Schr\"{o}dinger cat states) \cite{DBRH96}.

For temperature $T$ we have 
\begin{equation}
\frac{\tau _{{\rm th }}}{\tau _{{\rm dec }}}=\frac{2\left| \alpha \right|
^{2}\tanh \left| \alpha \right| ^{2}}{\bar{n}\left(\omega\right)}
\left\{ \left[ \bar{n}\left( \omega \right) +1\right] \left|
\alpha \right| ^{2}\tanh \left| \alpha \right| ^{2}+\bar{n}\left(
\omega \right) \left( \left| \alpha \right| ^{2}\tanh \left| \alpha \right|
^{2}+1\right) \right\}  ,
\label{ratthdec}
\end{equation}
which stresses the fact that decoherence is even faster than thermalization.
In the same way 
\begin{equation}
\frac{\tau _{{\rm dis }}}{\tau _{{\rm dec }}}=2
\left\{ \left[ \bar{n}\left( \omega \right) +1\right] \left|
\alpha \right| ^{2}\tanh \left| \alpha \right| ^{2}+\bar{n}\left(
\omega \right) \left( \left| \alpha \right| ^{2}\tanh \left| \alpha \right|
^{2}+1\right) \right\}  ,
\label{ratdisdec}
\end{equation}
where $\left| \alpha \right| $ plays its crucial role, and for large
temperatures we obtain the expected behavior of linear increasing of
decoherence rate with respect to temperature, in complete agreement with
expression (20) of \cite{JK98} and also with \cite{KB92}, but obtained in a
very different framework. To authors knowledge, it is the first time general
timescale relations as Eqs.\ (\ref{ratthdec}) and (\ref{ratdisdec}) appears
for this important example.

\emph{Master equation -- a quantitative indicative of separability}: 
The time scale of the reservoir can be straightforwardly evaluated by using 
the expansion Eq. \eqref{expandLinEnt} now for the linear entropy of the reservoir 
$\delta_2(t) = 1 - \Tr_2\rho_2(t)^2$. If we have 
$\Tr_2[\rho_2^2(0)-\rho_2^2(t)]/\Tr_2\rho_2^2(0) \approx 0$, we can consider 
that $\rho_2(t) \approx \rho_2(0)$. However, up to second order we get 
\be
1-\frac{\Tr_2 \rho^2_2 (t)}{\Tr_2 \rho^2_2 (0)} = t^{2}\sum_{k=1}^{N}
\gamma_{k}^2 \frac{\sin ^{2}\left( \Delta _{k}t\right) }
{\left( \Delta _{k}t\right) ^{2}}\left\{ 4\left( 2\left\langle a^{\dagger}a\right\rangle +1\right) 
\left[ \tanh \left( \frac{\beta \Omega _{k}}{2}\right) -1\right] 
+ 2\left\langle a^{\dagger }a\right\rangle -\left\langle a\right\rangle 
\left\langle a^{\dagger }\right\rangle \tanh \left( \frac{\beta \Omega _{k}}{2}\right) \right\} . 
\ee
Using the same procedure as before one can obtain the characteristic 
coherence change for the reservoir as 
\be
\tau_{2,\text{dec}}^{-1} = \frac{g(\omega)\gamma^2(\omega)}{2\bar n(\omega)+1}
\left| 2\left\langle a^{\dagger }a\right\rangle \left[2 \bar n (\omega)+ 1 \right] 
-8\bar n (\omega) \left(2 \langle a^{\dagger }a\rangle+ 1 \right)- \langle a\rangle 
\langle a^{\dagger }\rangle \right|. 
\label{tauDec2}
\ee

A reasonable condition for the separability of the timescales of systems 1 and 2 is 
\be \tau _{2,\text{dec}} \gg \omega^{-1}\,, \label{eq:tau} \ee 
where $\omega^{-1}$ is the characteristic time of the free evolution of system 1. 
The worst case we have is when the temperature is zero and therefore 
$\bar n(\omega)=0$. In this regime we can guarantee the use of the approximation 
$\rho_2(t) \approx \rho_2(0)$ if we have 
\be
g(\omega) \gamma^2 (\omega)\left\{2 \langle a^{\dagger }a\rangle 
- \langle a\rangle \langle a^{\dagger }\rangle \right\} \ll \omega.
\ee
Just to have some magnitudes to work with, we took from Ref. \cite{paris} the values $
\langle a^{\dagger }a\rangle \approx 9.5$, $\bar n(\omega) = .05$,
$\omega/2 \pi = 51 \, \mathrm{GHz}$, and $\tau _{\mathrm{dis}} = 160 \, \mathrm{\mu s}$,
and obtain $\tau _{2,\mathrm{dec}} \approx 13 \, \mathrm{\mu s} \gg \omega ^{-1}
\approx 20 \, \mathrm{ps}$, as one would expect.

If we want to have a condition which allows for the determination 
of a cutoff temperature in terms of dynamical ingredients and initial 
conditions we should use the condition \eqref{eq:tau} in Eq. \eqref{tauDec2}. 

To summarize we have provided a quantitative support for the factorization 
assumption between the system of interest and the reservoir density matrices 
which is widely used in the decoherence literature. Moreover we give analytical 
expressions for typical correlation and relaxation timescales which depend only 
on general ingredients such as the Hamiltonian and the initial state.

M. C. Nemes acknowledges the support of CNPq. J. G. Peixoto de Faria
thanks to PROPP--UESC for financial support under grants 220.1300.324.

\end{document}